# Cities of Knowledge and Big Science in Developing Countries: Luxury or Investment? The GCLSI Case.

Víctor M. Castaño[1], Leonardo Lomelí-Vanegas[1], Giorgio Margaritondo[2], Vanessa Mejía-Casco[3], Claudio Pellegrini[4], Galileo Violini[5]

1.-Universidad Nacional Autónoma de México

2.- École Polytechnique Fédérale de Lausanne

3.- Centro Internacional de Física, Colombia

4.-SLAC National Accelerator Laboratory

5.- Centro Internacional de Física, Colombia/Universidad Tecnológica del Cibao Oriental, UTECO, Cotuí, Dominican Republic

**Abstract**

This article analyzes the feasibility of having a second synchrotron in Latin America, to be located, in principle, in a city within the Greater Caribbean region but open to all the continent. It is shown that an initiative of this sort is compatible with the economies of the region and would require a marginal increase of the current regional investment in science, which is broadly below that of other regions of the world, with peaks of low financing precisely in the Greater Caribbean. The project is not only feasible, but, beyond its purely scientific interest. it would have an impact for the development of cities in the region. The article is mainly focused to analyze this impact from the social, economic, and political point of view. It is shown that the return of the investment would have its break-even point long before the end of the expected lifetime of the infrastructure, and that through a system of smaller accelerators, that would be part of the same project, the benefit would not concentrate on the country hosting the facility. These smaller facilities could contribute to the national development as possible nuclei of cities of knowledge, project which belongs to the priority of some countries/cities of the region.

## I. CITIES OF KNOWLEDGE AND DEVELOPING COUNTRIES

The First World offers several examples of science hubs, which are not necessarily capitals. Some are large cities, like Boston, while others are smaller, such as Cambridge and Heidelberg. However, since the mid-20th century, a new dimension of this phenomenon has emerged. Entire science cities have been established, often as a result of political decisions, along with a process of urban development that encompassed basic infrastructure, housing development, trade and job opportunities.

The motivations behind these science hubs vary. In Finland, Espoo developed around a major technological industry, Nokia. In the United States, Silicon Valley thrived in a uniquely stimulating scientific environment. A particularly distinctive case is that of Trieste, in Italy. Its starting point was the Abdus Salam International Centre for Theoretical Physics (ICTP), established to break the scientific isolation of the Third World. This vision was driven by Abdus Salam, who later received the 1979 Nobel

Prize in Physics for research conducted at the Center, and Paolo Budinich, whose political strategy focused on the city's development.

In the context of this article, Tsukuba, Japan, holds particular relevance. Once a small rural town, it transformed into a hub for hundreds of scientific institutions. Among the first was a Big Science facility, KEK, a research particle accelerator.

It is also not rare that existing industrial parks have had a substantial growth following the establishment of a big science facility, as it happened in Spain when ALBA synchrotron was built.

These examples have also drawn the interest of developing countries. In Latin America, two notable cases stand out: Panama's successful experience following the transfer of the Canal under one of the Carter-Torrijos treaties,[1] and Ecuador's ambitious project, led by President Correa, creating Yachay's City of Knowledge in San Miguel de Urcuquí, a small town in the north of the country.

However, both initiatives were based on the creation of research centers and/or universities. In the case of Yachay, the original objective had the broad scope of being a planned city. It also included the goal of attracting venture capital for innovation and industrialization. Such a project was abandoned after the 2017 political change, and today only a university, Yachay Tech, remains[2]. In any case, in neither case did the development of large research facilities play a significant role. As a result, Big Science infrastructures in Latin America exist almost exclusively in the fields of astronomy and astrophysics.

When creating cities of knowledge, often synchrotron radiation sources play a major role. Tsukuba and Trieste (with the Elettra facility) are only two of very many cases, including for example Berlin, Hamburg, Stanford, Barcelona, Krakow, Pohang, Shanghai, Beijing and numerous others.

In these context two proposals emerged in 2021 in the Central American-Caribbean region. An academic group (GCLSI) proposed the creation of a synchrotron radiation facility within Greater Caribbean (GC), where GCLSI stands for Greater Caribbean Light Source Initiative,[3-4] while, in Dominican Republic, President Abinader announced plans for a 'DR Silicon Beach', a city of knowledge to be established in Puerto Plata, a major tourist destination of the country.[5] The potential economic impact of this plan has occasionally drawn attention from both the national press[6-7] and academic circles[8].

This article has a twofold purpose: to examine the financial feasibility of such projects and to analyze their interplay, with a particular focus on the economic impact of establishing light sources in cities within developing countries. Our geographical scope will be the GC region; however, the analysis of the interaction between these proposals, along with the current lack of a decision about the location for the synchrotron project, means that our reference to a potential city of knowledge associated to the project, though inspired by the Dominican initiative, could extend to wherever the synchrotron may ultimately be built or even to multiple sites if a network of smaller facilities is established.

## II. CAN DEVELOPING COUNTRIES AFFORD LARGE RESEARCH FACILITIES AND CITIES OF KNOWLEDGE?

Empirical evidence linking economic growth with knowledge accumulation primarily stems from studies on high- and medium-high-income countries.[9-10] However, this general observation can obscure the distinct contributions of scientific, technological, and entrepreneurial knowledge.[11] A particularly relevant question, especially for developing countries, is: what exactly constitutes "scientific knowledge"? Scientific progress in these regions is often assessed through metrics like the number of scientists per capita or publication-citation counts. Yet use of these indicators may overlook the critical role of scientific infrastructure, particularly in the domain of Big Science, in enhancing their novelty, value and impact.

The 2024 Nobel Prize in Economics was awarded "for … studies on how institutions form and affect prosperity." The relationship between institutions and prosperity extends beyond the legacy of political and economic structures introduced by colonial powers. The fact that the wealthiest 20 percent of countries are approximately 30 times richer than the poorest 20 percent cannot be solely attributed to social institutions. To narrow this gap between developing and advanced countries, it is essential the creation of new institutions that include science and technology sectors as integral components.

Big Science institutions are essential for sustainable development. However, in developing countries, large-scale scientific research facilities often seem unattainable. Scientific funding is limited, typically accounting for only a small fraction of their GDP.

Among Big Science initiatives, synchrotron radiation sources offer a distinct advantage: their applications benefit numerous scientific fields and have a transformative impact across diverse sectors of economic activity. Building scientific infrastructure like synchrotrons can create jobs, attract foreign investment, enhance quality of life, and promote long-term economic growth. Initial costs may be high, but the investment is well justified by their direct positive effects in critical areas such as health, agriculture, and technology, sectors that are directly tied to some of the greatest challenges of the 21st century, including climate change and emerging diseases, not to mention an indirect contribution to local/urban development.

To illustrate these considerations, we will examine, as a case study, the feasibility of GCLSI, as a second facility of this sort in Latin America, where there exists only one, SIRIUS, in Campinas, Brazil. This case study will enable us to explore how such a project, when integrated into a broader initiative like a city of science, could catalyze a development that extends beyond its scientific value.

Further, we will demonstrate how, through collaboration, developing countries in that region can establish a network of major research facilities that provide significant economic, social, and human capital benefits for the cities involved.

First, we outline the multidisciplinary advantages of a GC synchrotron laboratory, followed by a discussion of the associated costs and the funding contributions required from each participant country, expressed as a percentage of their GDP and compared to their current science expenditures. Finally, we examine the benefits and long-term gains that such a research facility can offer to these countries. Indirect benefits will come from the synergy with similar projects in other regions, and from it possibly being the core, or at least one of the main institutions, around which the host country can develop a city of knowledge.

Our aim is to make a comprehensive analysis, that allows to identify a number of immediate actions as essential for the advancing of the project.

They fall along three main lines: first, exploration of the non-scientific factors that can justify the participation in the GCLSI, irrespective of the country level of development; second, assessment of three critical impact areas: poverty alleviation, medium- to long-term economic growth, and implications for regional political and scientific diplomacy; and third understanding of how this kind of facility can play a role for specific programs of local development such as cities of knowledge, indeed a novel approach in this type of analysis.

## III. SYNCHROTRON RADIATION SOURCES AND THEIR POTENTIAL IMPACT ON DEVELOPING COUNTRIES

Synchrotron radiation sources, or simply synchrotrons, are unique facilities with distinctive characteristics. A notable sociological study highlights that synchrotrons are supported by tightly connected international scientific communities, where ongoing interactions and collaboration foster collective advancement.[12] Insights gained from research at one synchrotron are often enriched by the experiences and technical progress achieved at others, forming a strong network for knowledge transfer. This feature enables synchrotron facilities to bridge fundamental research with applications to critical global challenges. A striking example of this is the “SYNAPSE” collaboration[13], which brings together around twenty synchrotrons worldwide (about a third of all synchrotrons) to address a pivotal research question: mapping the neural connections of the human brain. The scientific and societal impact of this initiative promises to be groundbreaking for the field of neurology.

Synchrotrons also represent a distinctive model of internationalization, diverging significantly from traditional approaches to scientific collaboration. Two successful multi-country examples illustrate this potential. At one end of the spectrum, in terms of scale and cost, is the European Synchrotron Radiation Facility (ESRF), an advanced laboratory built by a coalition of European countries and located in France. A smaller, more cost-effective, yet scientifically and politically impactful example, is SESAME, a Middle Eastern synchrotron collaboration, promoted by UNESCO and CERN, involving Egypt, Iran, Turkey, the Palestinian Authority, Israel, and others, situated in Jordan.

Big Science facilities primarily focused on fundamental research, such as CERN, Fermilab and KEK lead to technology transfer and spillover effects that also extend to pioneering technologies crucial for experimental success, and this generates substantial economic benefits.[14] Unlike these facilities, synchrotrons operate at the intersection of fundamental research and technological innovation, but this does not prevent them from achieving groundbreaking scientific discoveries. Notably, approximately two dozen

Nobel Prizes have been awarded for research involving synchrotron-based studies, the latest being the 2024 Chemistry Prize awarded to David Baker, who utilized findings from the Brazilian synchrotron SIRIUS to identify an enzyme critical to his AI-driven protein research.[15-16] However, synchrotrons' practical impact is further enhanced by economic and political drivers, expanding their mission to support industrial development and applied sciences. This unique role positions them as essential bridges between basic research and applied technology.

The wide range of their applications has a profound potential to impact the economy and make them valuable assets across diverse fields. Synchrotrons, for instance, enable analysis of materials essential for renewable energy, environmental monitoring, climate modeling, and the development of innovative materials and metamaterials, as well as the detailed study of ancient cultural heritage. They are particularly effective for studying the environmental effects of heavy metals, especially those from human activities. and their links to various diseases.[17] Additionally, their applications span ocean chemistry, agricultural productivity, health, medicine, and structural biology.

## IV. GCLSI, AN AMBITIOUS REGIONAL PROJECT

In this context, we analyze the economic and political feasibility of establishing a synchrotron in the GC region, with a vision to its contribution to a more localized development as well, that could be the ultimate goal of the establishment of a city of knowledge.

The economic feasibility is quantitatively assessed using two indicators: an absolute measure—representing the cost as a percentage of the countries' GDP—and a relative measure, which compares this cost, as a fraction of GDP, to the current allocation for STI. The political feasibility is evaluated based on the infrastructure's anticipated impact and projected return on investment (ROI), which must be substantial enough to warrant selecting such a facility as a developmental catalyst. Additionally, we consider intangible, yet crucial, aspects of political impact: the role of regional science diplomacy and that of creating a network of national hubs of science.

Our findings are unequivocal: the GCLSI is not only feasible but also represents a timely and promising opportunity. Applying the Romer model[18]—an effective tool for estimating economic growth and assessing public policy and long-term investments—we conservatively estimate a benefit-to-cost ratio of 1.75.

This result is not surprising, for it aligns with ex post calculations for the DIAMOND synchrotron in the United Kingdom[19-20] and PETRA in Germany[21]. DIAMOND yearly contribution to UK gross GDP has been estimated as 10% of the total investment, but even more impressive is the evaluation of PETRA III outcome. DESY's investment in building and operating the facility has been 815 million euros, about 65 million per year and PETRA III has produced unique scientific insights and provided essential data for innovative developments, whose added value amounts to **more than 2.25 billion euros**, (ref: https://photon-science.desy.de/news__events/news__highlights/archive/archive_of_2023/new_study_desys_light_source_petra_iii_has_created_225_billion_euros_in_added_value/index_eng

.html) even without including in the analysis greater additional benefits result of complementary, structural effects. The economic and industrial environment of Germany and UK is obviously different from that of GC and this is reflected on our lower estimated return, but great complementary returns at level of local development are certainly to be expected.

Furthermore, we find that the investment could reach a break-even point if it contributes just 0.055% to regional GDP growth, a level that, based on the documented performance of similar facilities worldwide, is realistically achievable.

To develop these projections, it is essential to identify and quantify both the direct and indirect costs and benefits of the project. Our quantitative analysis excludes certain intangible or long-term future benefits. In particular, we do not attempt to assess the additional costs or benefits associated with embedding the facility within a broader city of knowledge. However, evidence from several cases, most notably Tsukuba, suggests that this effect can be significant. As a result, our estimates remain conservative while still demonstrating that the GCLSI is both a feasible objective and a valuable opportunity for GC countries.

Although the challenges are substantial, the potential rewards could be even greater. Given that GCLSI activities will span 25–30 years, all cost and benefit estimates are calculated in 2024 US dollars (USD). This approach allows us to determine the investment's break-even point, which. based on our conservative cost assumptions, is projected to occur roughly one-third of the way through the facility's operational lifespan. Additionally, many of the benefits will be infrastructural, extending well beyond the facility's active years. A historical parallel can be drawn with the development of Boston as a scientific hub, which traces its roots to the foundation of Harvard in 1636 and to the Morrill Act of 1862, which granted MIT land-grant university status.[22]

We recognize some inherent limitations in this approach. Certain direct benefits will only start to materialize once the facility is fully operational, which would warrant applying at least a 5-year discount factor.[23] A reasonable estimate of this adjustment reduces the benefit-to-cost ratio by approximately 10–15%. However, if we consider this offset by the extended infrastructural benefits mentioned, among which a special role must be associated with the possible establishment of a regional city of knowledge, then the ROI estimate of 1.75 can be retained as reliable, and possibly conservative.

## V. A GLOBAL PENALIZING DISPARITY

While political leaders in developing countries often highlight the importance of science in their campaign rhetoric, promising to allocate the "mythical" 1% of GDP to Science, Technology, and Innovation (STI), as recommended in periodic UN surveys, this commitment frequently wanes in the face of pressing daily challenges. The high volume of social demands, coupled with fiscal constraints often driven by austerity policies, results in relatively low spending on science and technology. Consequently, Big Science projects are often sidelined.

Scientific research is inherently costly, and as it progresses, the expenses for infrastructure, equipment, and maintenance tend to increase. Ambitious projects necessitate strong political commitment at the highest levels, ideally involving both government and private sector stakeholders who recognize the long-term multiplier effect of investments in science and technology, particularly in Big Science. This shared vision is crucial for elevating such projects to state priorities, overcoming resistance, and enduring political shifts.

Funding may become more accessible when these projects align with the scientific priorities of advanced countries or global partnerships can enhance research infrastructure, as the South African Square Kilometer Array (SKA) project illustrates.[24]

In Latin America and the Caribbean, challenges are further exacerbated by structural inequalities and power imbalances, which can lead to conflicting interests among public and private actors, adding complexity to decision-making processes. In comparison to the relative abundance of astrophysics facilities, where geographical advantages have facilitated the establishment and operation of significant observatories, such as the Chacaltaya Laboratory in Bolivia, supported by Japan in the mid-20th century, synchrotron radiation facilities are notably scarce in the Global South: Africa has none, while Latin America and the Caribbean have only SIRIUS. This gap is particularly significant given the diverse applications of synchrotron technology that are directly relevant to the needs of African and Latin American countries.

This disparity has drawn the attention of the global scientific community. For instance, the African Light Source (AfLS) project, proposed decades ago, has gained momentum, with Ghana becoming the first country to offer political support.[25] Similarly, there have been synchrotron proposals in Latin America over the years, including initiatives in Colombia[25] and Mexico[27].

The GCLSI has accelerated this process over the past five years generating renewed interest[4-28-29] due to its regional focus and fostering interregional collaborations. Despite this growing enthusiasm, the initiative's potential impact may not yet be sufficient to secure government endorsement or broad public support, especially if it is primarily viewed as a scientific endeavor focused on physics and chemistry. The broader socioeconomic benefits risk being overlooked unless they are communicated effectively.

## VI. OBJECTIVES OF THE PRESENT ANALYSIS

The high costs associated with establishing such infrastructures, often amounting to hundreds of millions of dollars, pose a considerable challenge.

It is obvious that, at an individual level, the cost is unaffordable for most countries in the GC region. However, GCLSI is a regional project, and the cost would be shared. Nevertheless, even for the national quota, are these investments realistic for countries grappling with pressing issues like healthcare, food security, and education? In several countries, the problem of unschooled childhood remains stagnant, affecting them disproportionately compared to high-income countries.[30] Can countries with deep social inequalities come together to prioritize such a regional project?

While a commitment to ensuring that no participating country is left behind might foster collaboration, it remains uncertain whether this sentiment alone will attract the interest of less-developed countries. Even if, as we will demonstrate, their financial burden would be relatively modest, the question persists. This requires a diffused dissemination of the expected benefits at regional and national level.

So, what kind of return can be expected from an initiative like the GCLSI? Can the requested investment be justified by weighing its potential benefits against its costs? Is a process like the one that allowed KEK to make Tsukuba grow to its present level, possible also in a country of the Global South, or only in the First World? Will the outcome depend on the differences between the two cases?

Synchrotron facilities hold significant promise for poverty alleviation—an enduring challenge. In 1850, poverty eradication was urged as a universal duty by Victor Hugo before the French parliament[31], but two centuries later, the UN's 2030 Agenda still enshrines this goal as the first priority of the Sustainable Development Goals. Latin America has put on place numerous policies and actions aimed at poverty reduction. What did they achieve?

While the COVID-19 pandemic temporarily disrupted poverty trends, its impact began to recede in 2023. However, a longer-term perspective does not offer much optimism.[32] For example, between 2008 and 2022, global extreme poverty increased from 9.1% to 11.2%. A slight improvement in rural areas was offset by a doubling of poverty rates in urban regions, despite the fact that, during the same period, global GDP grew consistently at an average rate of 2–4%, although with some regional variations. This provides an additional reason to call for non-conventional mechanisms to promote urban economic development.

Such a negative correlation, along with a mere 1% decrease in unschooled childhood over the last decade [32], further challenges the justification for underfunding research in developing countries, namely the high opportunity cost associated with addressing more immediate needs. It casts doubts on the effectiveness of traditional anti-poverty measures within a reasonable timeframe. Additionally, the recent World Bank recommendations suggesting that developing countries focus on labor market modernization through skill-building and education[33] appear as well to have limited real effectiveness. This questions the premise of the approach, according to which the interaction and relationship between skills and technology are fundamental to explain the observed differences in productivity and income across countries, thereby reducing the issue to merely educational interventions.[34]

The lack of impact on income, combined with a decline in productivity across Latin America for over a decade, reported by CEPAL,[35] despite increased access to higher education, calls for a reevaluation of this entire approach.

The underlying philosophy that developing countries can overcome inequalities through targeted interventions fails to challenge their structural linkage with service-based economic models where science is often seen as a luxurious, neocolonialist component catering to the needs of advanced countries. Ultimately, the idea that the gap will close solely through the enrichment of poorer countries has been challenged by the conclusions of the 2024 Nobel Prizes in Economics.

There is substantial evidence that existing synchrotron facilities have brought significant economic benefits to their host countries, regardless of initial economic status. These benefits include job creation (beginning in the design and construction phases), attracting foreign investment, and strengthening industrial capacity, all of which contribute to an improved quality of life and can be fertilizers for the establishment of hubs of science not necessarily related to light sources.

In Spain, the ALBA synchrotron’s impact has been evaluated across three stages. Initial estimates indicated that construction costs would be offset by positive effects on production and employment even during the building phase, with a projected benefit-to-cost ratio exceeding 1.5 over 25 years.[36] The outcomes confirmed these projections[37], and similar forecasts have been made for the upcoming ALBA II upgrade[38]. Notably, ALBA attracted approximately 100 new businesses to the region, including several international firms.[39] A comparable economic boost has been observed at the Swiss Light Source (SLS) facility in Switzerland, particularly benefiting its vital pharmaceutical industry.[4’] Similarly, in Brazil, the pharmaceutical sector has experienced notable growth, partly due to its synchrotron facility.[41] In Japan, strong industrial partnerships have developed around the SPring-8 synchrotron.[42]

These concrete examples from countries across three continents provide persuasive evidence that the GCLSI can achieve similar success in the GC region.

The financial feasibility of such an investment warrants careful evaluation. We will demonstrate quantitatively that the cost of building and operating a synchrotron facility is within the financial reach of Caribbean economies. Furthermore, this investment could substantially increase regional research funding, which currently does not exceed a modest 0.3% of GDP in any country within the region. Raising this funding level could help the region narrow its research spending gap with Europe and Asia, or at least reach levels seen in more advanced African countries such as Tunisia, Morocco, South Africa, and Kenya.

## VII. COST ANALYSIS

A synchrotron facility involves two main cost components: construction and operational expenses, with operational costs also including occasional expenses for building new beamlines. Construction costs for facilities such as SIRIUS in Brazil, ALBA in Spain, and NSRRC in Taiwan have ranged between $200 million and $300 million USD, showing relative stability in the decade prior to the pandemic. A detailed 2023 cost estimate for a similar project in Iran[43], using a USD-to-Euro exchange rate of 1.08, also indicated a total cost of $224 million USD, covering buildings, the accelerator facility, and seven beamlines.

Land costs must also be considered. However, estimating their impact is complex—not only because the host country of GCLSI could donate the necessary land, but also dueto the potential increase in land value, particularly if the project is linked to the development of a city of knowledge. A relevant precedent is the case of Hidalgo, where the state offered land for the proposed Mexican synchrotron project.[44]

For the Iranian project, the land value was estimated to be approximately $7 million USD, bringing the total estimated cost of the project to around $230 million USD. To remain cautious, however, our cost-benefit analysis will use a higher estimate of $300 million USD, to account for potentially higher costs due to the relatively limited industrial infrastructure in the Caribbean compared to Iran and other synchrotron sites.

In addition to construction costs, operational expenses are estimated at approximately $25 million USD per year, based on similar facilities. With a projected facility lifespan of 25 years, including 5 years for construction and 20 years for operation, total operational costs are approximately $0.5 billion USD. Therefore, the total estimated cost for the GCLSI project is around $0.8 billion USD (in 2024 USD).

It is worth noting that this operational cost estimate is conservative for several reasons. It excludes potential energy savings achievable by adopting a sustainable energy model similar to SESAME, which operates entirely on solar power—a model particularly beneficial in promoting renewable energy for large facilities in the region. Additionally, this estimate is higher than typical current costs for two reasons: it includes the possibility of establishing a network of smaller accelerators, which we will discuss as a potential strategy to ensure broader regional participation, and it aims to make our use of 2024 USD value insensitive to the possibility of an annual rise of the operating cost in the upper part of the interval 3-6%.

Funding for the GCLSI could be sourced through international cooperation programs or development banks. For what concerns the former, the present geopolitical situation can make it uncertain. However, for the latter, the GC region has access to financial institutions like the World Bank, the Inter-American Development Bank, and the Central American Bank for Economic Integration (CABEI). During the COVID-19 pandemic, CABEI provided a $400 million USD credit facility as part of an economic recovery program[45]. A fraction of this amount would be sufficient as seed funding for the GCLSI, helping to attract additional international backing.

Operational funding for the GCLSI would likely be shared among participating countries. Initially they will be limited to those within the GC region. However, it is foreseeable that additional Latin American countries—or even non-regional countries—could join as Associate Members, following a model similar to CERN's. The recent adhesion to SESAME as associate members of Germany and Italy, confirms that it is a realistic possibility. Under CERN's approach, contributions are scaled proportionally to each country's GDP, up to a set limit, offering a potential framework for both governance and financing.

Table 1 outlines projected contributions for operational funding, assuming participation from GC countries and Ecuador, a country with some representation in GCLSI.

As with CERN, no single country would be expected to contribute more than a specified cap. The table presents results under two scenarios, with maximum contribution caps of 25% and 30%, respectively. In the unlikely scenario that participating countries would finance the entire construction cost, the average annual contribution over 25 years would increase by 28%.

Table 1. Countries contributions to operational costs

| Country | 30% cap | 25% cap |
|---|---|---|
| | **Contribution amount (MUS$)** | **Contribution amount (MUS$)** |
| Antigua | 0.022 | 0.036 |
| Barbados | 0.068 | 0.11 |
| Bahamas | 0.154 | 0.248 |
| Belize | 0.034 | 0.054 |
| Colombia | 4.119 | 6.621 |
| Costa Rica | 0.83 | 1.334 |
| Cuba | 0.168 | 0.27 |
| Dominica | 0.007 | 0.012 |
| Dominican Republic | 1.358 | 2.183 |
| Ecuador | 1.375 | 2.21 |
| Grenada | 0.015 | 0.024 |
| Guatemala | 1.135 | 1.825 |
| Honduras | 0.379 | 0.609 |
| Haiti | 0.224 | 0.36 |
| Jamaica | 0.204 | 0.329 |
| Saint Kitts and Nevis | 0.012 | 0.019 |
| Saint Lucia | 0.027 | 0.044 |
| Mexico | 7.5 | 7.143 |
| Nicaragua | 0.187 | 0.301 |
| Panama | 0.915 | 1.47 |
| El Salvador | 0.388 | 0.624 |
| Trinidad and Tobago | 0.359 | 0.578 |
| Saint Vincent and the Grenadines | 0.012 | 0.019 |
| Venezuela* | 1.339 | 2.152 |

Given the projected costs and proposed funding distribution, a key question remains: Are the expenses of establishing and operating the GCLSI compatible with the economic conditions and science policies of the participating countries?

To address this question, it is essential to review the available data—albeit limited—on STI funding across the region. For context, we refer to the most recent figures on Research and Development expenditure as a percentage of GDP, sourced from the World Bank[46] and the UNESCO Science Report[47]. Unfortunately, these data cover only a few countries and are often outdated.

Table 2 provides a summary of these figures.

Table 2 Available most recent STI funding data for GC countries.

| Country | STI funding as GDP % | Year | Amount in USM$ (2023 GDP data unless differently specified) |
|---|---|---|---|
| Colombia | 0.29 | 2020 | 783.44 |

| Costa Rica | 0.28 | 2021 | 181.87 |
|---|---|---|---|
| Cuba | 0.31 | 2021 | 332.79 |
| El Salvador | 0.16 | 2021 | 46.91 |
| Guatemala | 0.06 | 2021 | 51.59 |
| Honduras | 0.06 | 2019 | 17.19 |
| Jamaica | 0.06 | 2002 | 2.42 (2020 GDP) |
| Mexico | 0.27 | 2022 | 4544 |
| Nicaragua | 0.11 | 2015 | 17.27 (2020 GDP) |
| Panama | 0.18 | 2022 | 137.74 |
| Saint Lucia | 0.27 | 1999 | 6.20 (2020 GDP) |
| Saint Vincent and the Grenadines | 0.21 | 2002 | 1.89 (2022 GDP) |
| Trinidad and Tobago | 0.19 | 2016 | 40.65 (2020 GDP) |
| Venezuela | 0.69 | 2016 | 302.15 |

These data show that each country's contribution to the GCLSI project would represent only 1-2% of its current STI funding. Although this figure may seem substantial, it is important to recognize that STI investment across the GC already falls short of the Latin American average of 0.62%, with Brazil's 1.3% significantly boosting that regional average. Thus, the financial and strategic goals of establishing a synchrotron laboratory in the GC remain within reach.

The 2019 World Bank report emphasized that developing countries need to allocate approximately 4.5% of GDP to infrastructure to achieve their infrastructure-related sustainable development goals.[44] Various insights from this report could influence decisions about the type of synchrotron facility to construct. For instance, access to healthcare in the region suffers from economic and social disparities, which extend beyond hospital resources to include advanced medical technologies. Synchrotrons, particularly through X-ray crystallography, are essential for studying proteins associated with diseases like cancer[48] and diabetes[49]. They enable advanced X-ray microscopy, offering high precision in investigating brain functions and neurodegenerative diseases. Additionally, accelerators aid in producing radioisotopes for medical treatments[50], highlighting a synchrotron's potential to enhance healthcare capabilities significantly.

Numerous critical areas politically justify funding a synchrotron. One example is research related to climate change and sustainability, where the expertise and human resources developed at the synchrotron would be invaluable for Caribbean countries engaged in international initiatives to create action plans and collaborations aimed at monitoring and mitigating climate damage.[51] Another significant example is vaccine development. A critical step in developing the COVID-19 vaccine was determining the virus's structure. The global network of synchrotron facilities, particularly Brazil's SIRIUS, was rapidly mobilized at the start of the pandemic, successfully identifying the virus structure within weeks, which accelerated the vaccine development process.[52]

Mining is another promising sector[53], particularly with emerging opportunities in rare earth mineral extraction[54], which could greatly benefit from the advanced research capabilities of a synchrotron facility. In addition, advanced research could support a more environmentally sustainable mining industry, helping to address longstanding challenges.[55]

Another essential consideration is the macroeconomic landscape and its indirect impact on poverty reduction. The economic structures of many countries in the region remain relatively homogenous, with mining and agriculture continuing to dominate, much like two centuries ago[56]. However, sectors such as tourism, services, informal internal trade, and remittances have also become significant contributors to regional economies. Unfortunately, these sectors are highly volatile and, especially in the case of services and remittances, are often beyond national control.

To these considerations adds the aspect of the development of one or more regional cities of knowledge, in a region where science, that for historical reasons did not participate of the industrial revolution, may represent a unique accelerator of development.

## VIII. BENEFIT ANALYSIS

We will now continue our analysis of the investments and benefits associated with the GCLSI. Studies on the return on investment (ROI) for Big Science initiatives typically focus on advanced countries[57-58], while the context in developing countries remains less explored. These countries face unique challenges, which broaden the scope of ROI to include specific intangible benefits beyond direct financial returns, such as improvements in higher education, the establishment of local doctoral programs, and the reduction of brain drain[59-60].

How do the direct benefits of the GCLSI compare to its investments? We estimate its direct costs at $0.8 billion USD over 25 years. The Romer method also requires an assessment of indirect costs and a discussion of model sensitivity[18]. Regarding indirect costs, we expect these to be minimal, with the primary concern being potential impacts on energy infrastructure due to increased demand, although as we noted, solar energy could power such facilities, as for SESAME.

In terms of sensitivity, the project's success depends largely on the political commitment of regional governments and their ability to address existing discrepancies in light of a paradigmatic shift in international relations and to obtain internal social endorsement, which necessitates a non-partisan approach to gain widespread support.

A straightforward way to quantify the benefits is by using a reverse Romer method to address a key question: What portion of annual regional GDP growth must the GCLSI project contribute in order to achieve cost-benefit equilibrium?

The total GDP of the region is approximately $2.9 trillion USD. Assuming an annual growth rate of 2.5%—in line with the 2%-4% growth range observed over the past two decades—this equates to an annual increase of roughly $72.5 billion USD. We conservatively consider the GCLSI's expected impact on growth only during its 20-year operational period. Given total GCLSI costs of $0.8 billion USD, this breaks down to approximately $40 million USD per year.

To reach cost-benefit equilibrium, the GCLSI would need to generate an equivalent yearly revenue, which represents just about 0.055% of the projected annual GDP increase.

Is this cost-benefit equilibrium level reasonable and realistic? In fact, this target appears highly achievable based on historical data. For example, the British synchrotron DIAMOND demonstrated impressive results: between 2007 and 2019, with an investment of £1.2 billion, it generated an estimated impact of £1.8 billion by 2021[12], with an additional £0.8 billion when the estimate was extended over the subsequent three years[20]. Similarly, an impact study for Germany's PETRA facility showed comparable results.[21]

To estimate the GCLSI's direct contribution to GDP growth, we can examine data from other synchrotron facilities. For instance, the Canadian synchrotron contributed $33 million USD to Saskatchewan's GDP and $90 million USD to Canada's overall GDP. The Australian synchrotron, which dedicates at least 20% of its operational time to industrial applications, generates annual benefits of approximately $130 million USD. [27]

Given these figures—and acknowledging the different economic contexts—we conservatively estimate that the GCLSI will generate at least $70 million USD in benefits per year. This projection not only comfortably surpasses the cost-benefit equilibrium but also represents a benefit-to-cost ratio of 1.75, with an anticipated break-even point around the seventh year of operation.

## VIII. ADDITIONAL BENEFITS

Quantifying the return on investment (ROI) for major scientific facilities requires a nuanced approach that extends beyond simple GDP impact analysis. Applying the Romer model effectively entails considering various additional factors[61]. Rather than attempting to quantify the direct scientific benefits—such as growing the scientific community and producing impactful research and the possible establishment of big science hubs—this assessment will focus on technological advancements, including patents and human capital development.

One of the intrinsic challenges in evaluating technological impacts is their long-term nature, as secondary effects can often surpass the initial outcomes over time.[62] This consideration is particularly relevant when assessing certain indirect benefits, which necessarily are underestimated.

An important consideration when planning Big Science facilities, such as raised in the context of South Africa's Square Kilometer Array (SKA)[24], but of broader scope, is the need for a scientific community capable of fully utilizing the facility. Currently, only Mexico satisfies this requirement. However, in this regard, we refer to the SESAME experience that implemented a robust training program during the construction phase to build local expertise. This approach could be a valuable model for the GCLSI as well.

Advanced training and educational programs are critical for building human capital, but they also entail significant costs. Big Science initiatives can support the development of sustainable national and regional PhD programs, which are crucial in Latin America,[63] to reduce the need for students to study abroad and mitigate the costs associated with brain drain, as many PhDs who study overseas do not return. By establishing local PhD programs, it is reasonable to estimate that the cost savings could be comparable to the GCLSI's annual operating expenses.

Brain drain costs can be estimated by considering two main factors: training costs, which vary significantly between low-income and advanced countries[30] (with up to a two-order-of-magnitude difference), and the value of skills, which also depends on the economic context in which they are applied. Approximately 35 years ago, brain drain costs were estimated at $1 million per person, now adjusted to around $2 million[59-60]. Other estimates, although based on different criteria, suggest lower values, but still amount to hundreds of thousands per individual. Using the highest estimates, the GCLSI would only need to retain 15 researchers per year to achieve cost-benefit equilibrium. More conservatively, reducing brain drain alone could realistically deliver an annual economic benefit of at least $100-200 million.

Beyond financial metrics, reducing brain drain yields social and academic benefits. For example, Italy's Elettra facility, part of Trieste's hub of science, allowed researchers in the Trieste area to conduct world-class research locally. The benefit was not limited to Italy. It was extended to neighboring countries and fostered international cooperation and knowledge exchange.

Another benefit, challenging to quantify but significant, is the growth of the scientific community—not merely in size but in scope. Large-scale infrastructure projects tend to foster the creation of additional institutions and allow the establishment of cities of science. Trieste's ICTP has inspired the establishment of associated institutions like TWAS, SISSA, and ICGEB and the creation of a synchrotron, Elettra. Similarly, Brazil's synchrotron facility is part of CNPEM, which oversees three other national laboratories (Biosciences, Biorenewables, and Nanotechnology) as well as an undergraduate School of Science. Moreover, the need of managing big data, for projects like SYNAPSE contributes to participate in big data centers and develop expertise in intensive computer and artificial intelligence applications.

A strategic challenge lies in regional funding allocation, particularly for countries that will participate in but will not host the facility. Countries such as Colombia and Mexico may encounter this issue (see Table 1). While international collaboration offers many advantages, it also brings complexities tied to national interests, as seen in other synchrotron projects. This may be especially relevant in the GC region.

A well-designed governance structure is essential to mitigate these challenges. Establishing a network of smaller, cost-effective accelerators—funded by regional contributions and managed under a unified governance framework—could help address potential tensions. Our $25 million estimate includes the operative costs for this setup, although these elements are not directly reflected in our benefit calculations. However, despite the regional funding, a byproduct could be the establishment of national hubs of science that would guarantee a national return to the regional investment.

Balancing these political and logistical factors—keeping in mind the SKA lesson that quality must take precedence over quantity[24]—could help align participating countries toward a successful regional initiative.

Is a Light Source system of this kind feasible and sustainable? Given the military expenditures[64] and debt levels of most contributing countries, with debt ratios generally around 60% of GDP (except in Guatemala)[65], this endeavor appears manageable. In this context one can observe that the total GDP of only Central America and Caribbean Islands is of the same order of magnitude of that of Sweeden (593. 27 US in 2023). This country not only hosts a fourth-generation national synchrotron, located in the historically relevant, yet small city (82,000 inhabitants) of Lund, which also has become an important hub for science and technology in the region, but has collaborated also to the building of the Polish one.

However, this is not enough, and the long-term commitment of participating states is essential for the project' sustainability, which remains a crucial sensitivity factor in our estimates, as the Yachay case confirms[2].

Balancing the investment in Big Science with these strategic considerations is essential for enhancing regional collaboration, fostering broad participation, and maximizing benefits[66]. The impact of the GCLSI project should also take into account two additional factors:

1. **Intangible Cultural Value**: Public willingness to pay for the cultural and scientific benefits of such facilities is crucial, and the linkage to a project of City(ies) of Science might help. However, quantifying this value can be challenging, as it correlates with education levels and the public's understanding of the project—factors that can be variable in a region where the perception of science may face obstacles.[67]
2. **Third-Party Services**: Synchrotron facilities can generate additional revenue and promote international cooperation by offering services to external entities. For instance, studies have highlighted the innovation potential of synchrotrons, including the European Synchrotron Radiation Facility's (ESRF) adaptations to meet industry needs.[68-69]

In Latin America and the GC, evaluating the impact of industrialization—particularly in high-tech sectors—introduces additional complexities. In the 21st century, Latin America cannot yet be the potter land celebrated by Neruda.[70] The region has yet to fully transition from a natural-resource-based economy to attain true industrial independence through high-tech development. Brazil's 1999 decision to establish a synchrotron facility, culminating in SIRIUS, the world's second fourth-generation synchrotron (85% locally constructed), not only strengthened Brazil's scientific capabilities but also catalyzed industrial growth, notably in the burgeoning pharmaceutical sector.

Similarly, a synchrotron in the GC could stimulate regional industrial development, generating jobs and supporting poverty-alleviating sectors like healthcare. Among its potential benefits, two have a direct economic impact:

1. **Industrial Competence**: Developing regional expertise in technologies such as electronics and imaging could reduce dependence on high-tech imports, which represent up to 10% of GDP in some countries and even higher in some special imports in Mexico.
2. **Balance of Payments**: By promoting local high-tech industries, the region could shift from exporting raw materials to processing them domestically, thereby improving the trade balance and enhancing economic stability.

In summary, the GCLSI project holds significant promise for economic and industrial advancement across the GC and Latin America, reaching well beyond its immediate scientific contributions. Even a partial quantification of these benefits suggests a strong ROI, emphasizing its potential as a catalyst for regional growth and innovation.

## IX. POLITICAL CONSIDERATIONS

In summary, acknowledging the unsustainability of the current economic model and its limited effectiveness in reducing poverty, the additional benefits beyond those estimated in our simplified Romer model further validate the investment in GCLSI. These advantages suggest that the necessary positive decisions are not only feasible but also well-timed.

However, the economic impact is only one dimension of the project. Another key benefit is its political influence. Unlike past national initiatives, the GCLSI's regional focus has drawn considerable attention, underscoring the importance of cooperation among smaller countries in the region. This political aspect is crucial, as regional collaboration often demands setting aside local differences to work toward shared objectives. Moreover, it is timely in view of the uncertainties of the geopolitical situation. Science and its applications provide an ideal platform for fostering this unity. Notably, the GCLSI has already enhanced regional relations by hosting a multi-venue symposium across six countries[71], strengthening ties between Spanish- and English-speaking Caribbean nations.

Central America has historically been a pioneer in regional integration, with organizations such as ODECA (Organization of Central American States, later SICA), CETCAP (Central American Technical Council for Agricultural Production), CSUCA (Central American University Council), and CABEI. While a large scientific infrastructure was not initially, and part of these organizations' agendas, its relevance is growing, even amid the geopolitical challenges that may complicate its development.

The path to realizing the GCLSI project presents significant challenges. Advancing a project of this magnitude in Latin America and the Caribbean heavily relies on public higher education institutions and research centers, where most scientific research takes place. Universities play a pivotal role in fostering critical reflection on public investment priorities, sparking debate on which strategic areas should be prioritized in future policy agendas. It is also the responsibility of the fourth estate (the media) to facilitate this conversation and communicate the project's importance to society.

We understand that there may be doubts about the region's capacity to undertake such a project. However, Brazil's successful experience with synchrotron technology offers a compelling counterargument to these concerns, which mirror the doubts raised in Brazil four decades ago[27] regarding user demand, technological capabilities, and the potential negative impact on science funding. The successful implementation of projects like SESAME, ALBA, and SIRIUS, along with Brazil's growing investments in STI, demonstrate the feasibility of ambitious projects of this nature.

Furthermore, the economic theory of "creative destruction" [72] suggests that the establishment of GCLSI could catalyze transformative and disruptive changes in two key areas: science management across the region—driving collaboration between industry and academia, which is currently nearly nonexistent—and higher education, particularly in the sciences. The credentialism criticized by Schwartzmann[73] would be replaced by peer evaluation.

## X. CONCLUSIONS

The challenges and uncertainties outlined above should not deter support for a project whose regional necessity and feasibility have been clearly demonstrated. Our analysis identifies four immediate actions that are essential to advancing the GCLSI project, having in mind the role it may have to motivate the creation of an innovative city of knowledge in the GC Region.

1. **Official Prioritization**: Relevant governments must formally prioritize the GCLSI project. This commitment is crucial, as it ensures the project receives the attention and resources it requires. Once prioritized, governments can formalize high-level legal agreements concerning the facility's location, the establishment of a potential Compact Light Source network, and the governance of the scientific infrastructure. It is unrealistic to expect that a bottom-up initiative from GCLSI promoters could make decisions that belong to governments.
2. **Securing Support from Regional Development Banks**: Engaging regional development banks is vital, as it underscores not only the project scientific potential but also its economic and social benefits. Given the significant regional economic impact, the GCLSI is well-positioned to justify the investment required for its success.
3. **Enhanced Coordination with Similar Initiatives**: Strengthen coordination with parallel projects. While the GCLSI already has solid ties with AfLS, initiatives in Iran, Uzbekistan, and ongoing support for SESAME are other valuable connections. Collaboration among GCLSI/LAMISTAD, AfLS, SESAME, and other similar projects can leverage additional international support, create economies of scale in feasibility studies, training, and beamline operations, and attract broader institutional support from organizations such as ICTP and LAAAMP. Engaging UNESCO and other UN agencies is also critical, as synchrotron facilities should be recognized as key infrastructure for the scientific and technological advancement of developing countries within their mandates. A project vision grounded in the Global South could also resonate with broader UN organizations like UNDP and ECOSOC, and some indication for that came in a recent meeting at ICTP.

4. **Engagement with Existing Synchrotrons**: Actively seek collaboration with existing synchrotron facilities to secure beamline time for Global South projects. While this concept is not new, expanding it at an interregional level could help build larger user communities and foster South-South collaborations. Exploring the construction of lower-cost accelerators and assessing their outcomes would further emphasize the facilities' economic and social value for decision-makers.

The third action may be difficult in the current geopolitical situation. The goal of establishing the GCLSI under UNESCO's auspices[3-4] with a possible US support that takes advantage of US rejoining UNESCO and settling outstanding arrears, is extremely unlikely. Pressing global issues such as the future reconstruction of Gaza and Ukraine may be the medium-term priorities of the international agenda. However, these obstacles should not derail long-term regional policies. These challenges can be temporary and perhaps should not hinder pursuit of enduring initiatives, but we have shown that the Greater Caribbean do not need external support to build and operate a synchrotron, and that the reward recommends such a decision, independently of external conditions.

The fourth action introduces a broader principle for access to large scientific facilities[74], serving as a general guideline. Given the significant investment required for these facilities, often funded by First World countries, there is a moral imperative to ensure a wider and more equitable return. Supporting science in developing countries by facilitating access for their scientists—particularly young researchers—would help them acquire skills that directly benefit their home countries. This approach aligns with global goals of equitable and sustainable development.

While these steps will accelerate the progress of the GCLSI, significant challenges remain in securing full recognition of the essential role that Big Science can play in addressing regional structural issues such as poverty and inequality. Tackling these challenges is not only an economic necessity but a moral one. It is time for the universal right to scientific access to become a reality, not merely an aspiration in UN declarations.

If Big Science facilities continue to be concentrated in the Global North, we risk repeating the experience cited by the 2024 Nobel Prize in Economics, referring to the example of Nogales (Sonora and Arizona), divided by the US-Mexico border. "The American economic system offers residents north of the border greater opportunities to choose their education and profession, and they are part of the American political system, which grants them broad political rights. On the other hand, south of the border, the inhabitants live under different economic conditions, and the political system limits their ability to influence legislation." The result is a profound disparity between two similar regions and a City of Knowledge in the region could be an attractive initiative.

The primary agents capable of bringing the GCLSI to life are the region's political leaders. They can count on the region`s academic and research system, and on a growing international support.[75] This would allow them to rise to the occasion with a vision that transcends national boundaries.

**Will they accept this historic responsibility? If not now, when?**

Data availability

All data in this article is available, upon request

Inclusion and Ethics

All the applicable inclusion and ethics requirements have been included in this article

**References**

1- https://ciudaddelsaber.org/en
2- Machado, J., (2023), Ejecutivo ordena el cierre de la Ciudad del Conocimiento Yachay, Primicias, January 9, 2023, https://www.primicias.ec/noticias/sociedad/yachay-ciudad-conocimiento-cierre/
3- Violini, G.; Castaño, V.M.; Fuentes, J.; Gómez, P.; Medrano, G.; Posada, E.; Rudamas, C. (2021), A Synchrotron as Accelerator of Science Development in Central America and the Caribbean, https://arxiv.org/abs/2109.11979
4- Castaño, V., Fernández de Córdoba, P., Sans, J. A., Violini, G., (2024), Big science in Latin America: accelerate particles and progress, Nature, 627, 32, 2024
5- Abinader, L., (2021), Rendición de cuentas ante la Asamblea Nacional, 27 de Febrero 2021, https://minpre.gob.do/wp-content/uploads/2021/02/Discurso-Rendicion-de-Cuentas_Feb_2021.pdf
6- Violini, G., (2021), Silicon Beach no es utopía, acento.com, 31 de marzo 2021, https://acento.com.do/opinion/silicon-beach-no-es-utopia-8929082.html
7- De Jesús Salvador, W., Violini, G., (2022), Silicon Beach y el futuro de la economía dominicana, acento.com, 3 Abril, 2022, https://acento.com.do/opinion/silicon-beach-y-el-futuro-de-la-economia-dominicana-9049307.html
8- Violini, G., Ciencia en República Dominicana. (2025), Coyuntura y largo plazo acento.com, 16 de Febrero, 2025, https://acento.com.do/opinion/ciencia-en-republica-dominicana-coyuntura-y-largo-plazo-9456694.html
9- Aghion, P. et. al. (2021), El poder de la destrucción creativa. ¿Qué impulsa el crecimiento económico?, Barcelona, Ediciones Deusto
10- Moyo, C., Phiri, A. (2024), Knowledge creation and economic growth: the importance of basic research. Cogent Social Sciences, 10 (1), https://doi.org/10.1080/23311886.2024.2309714

11- Bacovic, M., Lipovina-Bozovic, M. (2010), Knowledge accumulation and Economic growth, in Economic Development, Tax system and Economic Distribution in the countries of Southern and Eastern Europe, 37-51, ASECU
12- Simoulin, V., (2016), The synchrotron generations, Communities and facilities at the crossroads between the national and the international, Revue française de sociologie, 57. 503, 2016
13- Stampfl, A. P. J., et al., (2023), SYNAPSE: an International Roadmap to Large Brain Imaging, Physics Reports 999, 1 (2023)
14- Scarrà, D., Pittaluga, A., (2022), The impact of technology transfer and knowledge spillover from Big Science: a literature review, Technovation, 116, 102165
15- Goverde, C. A., et al., (2024), Computational design of soluble and functional membrane protein analogues, Nature 631, 449-458
16- CNPEM News (2024), CNPEM research used to validate tool that led to the Nobel Prize in Chemistry, October, 10, 2024
17- Hu, H., et al., (2020), Synchrotron-based techniques for studying the environmental health effects of heavy metals: Current status and future perspectives, TrAC Trends in Analytical Chemistry 122, 115721
18- Romer, P., (2012), Advanced Macroeconomics (4th ed.), McGraw-Hill. New York
19- Technopolis Group, (2021), Socio-Economic Impact Report
20- Technopolis Group, (2024), Socio-Economic Impact Report
21- Kroll. H., et al., (2023), Impact-Studie Synchrotronstrahlungsquelle PETRA III im Kontext des Forschungs- und Innovationsökosystems DESY", DOI: 10.24406/publica-19299
22- *"MIT History | MIT Facts". libraries.mit.edu.*
23- Pearce, D., Groom, B., Hepburn C., Koundouri, Ph., (2003),Valuing the Future Recent advances in social discounting, World Economics, **4**:2, 121-141, 2003
24- Rüland, A.-L., (2022), Capacity-Building for Big Science in the Global South: Lessons Learned from the Square Kilometer Array, Journal of Science Policy & Governance, Vol. 20, Issue 3 (sciencepolicyjournal.org)
25- Wild, S., (2021), Plan for Africa's first synchrotron light source starts to crystallize, Nature News. https://www.nature.com/articles/d41586-021-02811-2
26- Gómez Moreno, B., (2014), Aceleradores para Colombia, Revista de la Academia Colombiana de Ciencias Exactas, Físicas y Naturales, 38, pag. 71-78, 2014
27- Del Rio, V., (2015) Estudio de Viabilidad para la Construcción de un Sincrotrón en México, UNAM
28- Latín American post staff, (2024), Advancing Latin America through Big Science: The Greater Caribbean Light Source, https://latinamericanpost.com/science-technology/advancing-latin-america-through-big-science-the-greater-caribbean-light-source/
29- RT (2024), https://xnwcoadmin.actualidad-rt.com/video/508984-diferencias-gastos-ciencia-paises-latinoamerica
30- UNESCO. (2024), Global Education Monitoring Report 2024/5: Leadership in education – Lead for learning. Paris, UNESCO
31- Hugo, V., (1850), Discours à l'Assemblée législative 1849-1851 La liberté de l'enseignement - Wikisource
32- CEPAL (2024), Statistical Yearbook for Latin America and the Caribbean 2023
33- Moroz, H., Viollaz, M., (2024), The Future of Work in Central America and the Dominican Republic. Washington, DC: World Bank
34- Sanchez-Paramo, C., De Ferranti, D., Perry, G. E., Gill, I., Guasch, L., Schady, N., Maloney, W. F., (2003), Closing the gap in education and technology, World Bank 35- Llinás, M., (2024), Ciencia, Tecnología e Innovación: Agendas regionales en América Latina y Caribe, UNESCO webinar
36- García-Montalvo, J., Raya Vilchez, J.M., (2005), Potenciant la nova economia a Catalunya: una anàlisi econòmica de la font de llum de sincrotró del Vallès (ALBA). Coneix. i Soc. Rev. d'Universitats, Recer. i Soc. la Inf. 32–59.
37- Raya, J.M., García-Montalvo, J.G., (2016), Anàlisi cost-benefici i d'impacte econòmic del sincrotró ALBA. Nota d'Economia 115–125
38- Raya, J.M., García-Montalvo, J.G., (2023), Impacto Económico y Social de ALBA II, Universitat Pompeu Fabra Report
39- Biscari, C,. (2024), private communication
40- See for example https://www.excelsusss.com/
41- Nascimento, A., (2021), Launch of the Manacá Beamline at Sirius: First Protein Crystallography Structures and New Opportunities for Pharmaceutical Development Using Synchrotrons, Synchrotron Radiation News, 34(5), 3–10. https://doi.org/10.1080/08940886.2021.1994310
42- Nonaka, T., et al. (2016), Toyota beamline (BL33XU) at SPring-8. Proceedings of the 12th International Conference on Synchrotron Radiation Instrumentation – Sri2015. https://doi.org/10.1063/1.4952866
43- Executive summary of ILSF (2023), ILSF-B-MN-0000-GRP-01-06E
44- Rozenberg, J. Fay, M., (2019), Beyond the Gap: How Countries Can Afford the Infrastructure They Need while Protecting the Planet (English). Washington, D.C. : World Bank Group
45- SICA (2020), SICA: BCIE emite bono por US$50 millones a cada país de la región para vacuna de la COVID-19 - Portal del SICA
46- Research and development expenditure (% of GDP) | Data (worldbank.org)
47- UNESCO. (2021), UNESCO science report: The race against time for smarter development. https://unesdoc.unesco.org/ark:/48223/pf0000377433
48- Vinciguerra, P., et al. (2020), Synchrotron Radiation-Based Techniques in Cancer Research: From Imaging to Therapy,. https://doi.org/10.3390/cancers12071913, Cancers, 12(7), 1913
49- Artyukhin, A. B., et al. (2021), Synchrotron-based X-ray Fluorescence Imaging of Pancreatic Islets in Diabetic Mice, Journal of Synchrotron Radiation, 28(5), 1234–1240.
50- Banerjee, S., et al. (2019), Synchrotron-Based Radioisotope Production for Medical Applications: Advancements and Future Prospects, Applied Radiation and Isotopes, 147, 170–176. https://doi.org/10.1016/j.apradiso.2019.02.013
51- Violini, G., (2024) https://www.researchgate.net/publication/388675168_GCLS_y_politicas_ambientales_y_de_educacion_superior
52- Ziegler, M. F., (2020), Brazilian synchrotron light source helps scientists look for COVID-19 drugs in first experiment, Agência FAPESP
53- De Jesús Hilario, I. C., (2024), Minería dominicana: visión y retos, Listín diario, September 11, 2024 https://listindiario.com/economia/industria/20240911/mineria-dominicana-vision-retos_825069.html
54- Jovine Rijo, F., (2024), Tierras raras: una oportunidad impostergable, Listin Diario, June 27, 2024 https://listindiario.com/la-republica/20240627/tierras-raras-oportunidad-impostergable_814466.html
55- Colón Rivera, J., Córdova Iturregui, F., Córdova Iturregui, J., (2014), El proyecto de explotación minera en Puerto Rico (1962-1968) Nacimiento de la conciencia ambiental moderna, Ediciones Huracán, Puerto Rico 56- Palacios, M., (1999), Independencia y subdesarrollo. Notas sobre los orígenes del liberalismo económico en Colombia, in Palacios, M., Parábola del liberalismo, Editorial Norma, Bogota
57- Florio, M., (2019), Investing in Science: Social Cost-Benefit Analysis of Research Infrastructures, MIT Press
58- Stankovski, M., Khotbehsara, F. A. P., (2024), What is the size of the global light- and neutron source research communities? — LINXS
59- Violini, G., (1991), La fuga de cerebros y sus implicaciones para políticas de Educación Superior, in Actas Reunión internacional de Reflexión sobre los nuevos roles de la educación superior a nivel mundial, López Ospina, G. Edt., UNESCO,
60- Violini, G. (1991), Some considerations on brain drain: the Colombian case, Discovery and Innovation, 3, 7
61- Florio, M., Sirtori, E., (2014), The Evaluation of Research Infrastructures: a Cost-Benefit Analysis, Milan European Economy Workshops, Working Paper 2014-10
62- Kurczynski, P., Milojević, S., (2020), Enabling discoveries: a review of 30 years of advanced technologies and instrumentation at the National Science Foundation. J. Astron. Telesc. Instrum. Syst. 6, 030901
63- Quevedo, F., Ordoñez, C., Violini, G., (2012), Propuesta de un programa de doctorados en Física y Matemáticas para las universidades del CSUCA CSUCA-ICTP
64- Stockholm International Peace Research Institute Database, accessed Septemebr2024, SIPRI Military Expenditure Database | SIPRI
65- World Bank (2024), Central government debt, total (% of GDP) | Data (worldbank.org)
66- Gutleber, J., (2021), Rethinking the Socio-economic Value of Big Science: Lessons from the FCC Study, in The Economics of Big Science, Beck, H., P., Panagiotis, Ch., Edts, Science Policy Reports, Springer
67- Álvarez, M., (2023), Percepción social de la ciencia y la tecnología en la República Dominicana. Principales resultados, https://www.researchgate.net/publication/373044782
68- Capria, E., (2024), The use of analytical research infrastructures to support industrial innovation, Materials week, Strategic R&I for the Value-Chains of the Future – Abstract S03_T07
69- Capria, E., et al., (2024), Adapting the European Synchrotron to Industry, Synchrotron Radiation News, 37,10,2024
70- Neruda, P., (1966), Foreword to La lira popular, in Neruda, P., Para nacer he nacido, Editorial Losada, 1977
71- LAMISTAD Symposium,(2023), Simposio LAMISTAD – Sincrotrón para el desarrollo sostenible (energesis.es)
72- Schumpeter, J., (1942), Capitalism, Socialism, and Democracy, Harper & Brothers
73- Schwartzman, S,, (2002), Higher education and the Demands of the New Economy in Latin America, Background paper for the World Bank's report on "Closing the Gap in Education andTechnology", Latin American and Caribbean Department, https://www.researchgate.net/publication/224771533_Higher_education_and_the_demands_of_the_new_economy_in_Latin_America
74- Feder, T., (2024), Research facilities strive for fair and efficient time allocation, Physics Today, September, 2024 https://pubs.aip.org/physicstoday/article/77/9/20/3309188/Research-facilities-strive-for-fair-and-efficient
75- Statement (2024), Joint ICTP-WE Heraeus Conference on Sustainability and Resilience Through Large Scale Infrastructures and Remote and Automated Experiments, ICTP, November 18-21